# Optical properties of Au-Hf thin films

*Hugh Littlehailes, William R Hendren, Robert M Bowman, Fumin Huang\**

*School of Mathematics and Physics, Queen's University Belfast, University Road, Belfast, BT7 1NN, United Kingdom*

\*Correspondence email: f.huang@qub.ac.uk

Abstract

The optical properties of thin films of intermetallic Au$_3$Hf were experimentally investigated for the first time, which display clear plasmonic properties in the optical and near infrared region with negative permittivity. In contrast to similar alloys, such as films of Au$_3$Zr, the films express more negative $\varepsilon'$ values and lower $\varepsilon''$ values across most of the wavelengths (370-1570 nm) investigated. The Au$_3$Hf films were fabricated by DC magnetron sputtering at a range of deposition temperatures, from room temperature to 415 °C, and annealed at different vacuum levels. The films mostly formed as a combination of Au$_3$Hf, Au$_2$Hf and Au$_4$Hf phases when deposited below 400 °C, and exclusively Au$_3$Hf phase at above 400 °C, indicating key conditions for isolating this phase. The films were stable when annealed at 10$^{-8}$ Torr, but when annealed again at 10$^{-6}$ Torr the films oxidised and changed into a mix of Au-Hf phases, suggesting resistance to oxidization may be an issue for un-encapsulated applications at elevated temperatures.

I.  Introduction

With the ability to confine optical energy below the diffraction limit [1-4], plasmonic materials have found applications in a broad range of areas, including subwavelength imaging [5], molecular sensing [6], superlensing [7], cancer therapy [8], and data storage [9], to name just a few. In most circumstances materials with large negative real part permittivity and small imaginary part permittivity (such as Ag and Au in the visible range) are favourable as they produce strong field enhancement and minimal loss. However, as the applications of plasmonics have extended to such a wide area, diverse applications have diverse demands on specific materials properties, which could be the optical, mechanical, chemical or thermal properties, dependent on the specifics of the applications. Some applications may require the materials to be catalytic, semiconductor or magnetic. As such, there is a continuous search for novel plasmonic materials [10]. Recently, the advent of plasmonics in data storage such as in heat-assisted magnetic recording (HAMR) and thermoelectrics require the development of plasmonic materials with strong mechanical strength and outstanding thermal stability endurable of elevated temperatures, which has prompted the search for novel refractory plasmonic materials [11]. Such investigations have looked into noble metal alloys, transition metal alloys, nitrides, germanides, silicides, intermetallics and conducting oxides [12-15]. Of these, intermetallic alloys stand out as an interesting avenue for exploration, with a vast number of alloy combinations available but still largely unexplored. Previous work in this field has shown promising results in the noble-metal binary alloys and alkali-noble compounds [14][16-20], such as $AuAl_2$, which expresses larger hot carrier generation rates than Au, and LiAu and KAu which have a damping frequency to plasma frequency ratios ($\gamma/\omega_p$) comparable to Au, and zero interband transitions below its plasma frequency, respectively.

In an earlier investigation [21], we demonstrate that refractory intermetallics alloy of $Au_3Zr$ exhibits notable plasmonic properties in the visible and near infrared regimes. Here in this investigation, we for the first time characterize the optical properties of Au-Hf alloys, with a prime focus on the 3:1 stoichiometry which has a melting temperature above 1500 °C [22]. Investigation into the Au-Hf system began with initial structural identification of some of the phases in the system by K. Schubert in 1960 [23]. Since then, there have been a number of subsequent studies that have sought to expand the number of known stable phases and refine the structural and thermodynamic characteristics of members of this system [24-30]. However, to date, no experimental data has been presented on the optical properties of Au-Hf system. Here we present for the first time the optical properties of $Au_3Hf$, demonstrating its plasmonic properties and thermal stability when annealed at elevated temperatures.

II. Materials and methods

Thin films of $Au_3Hf$ close to 100 nm thick were co-sputtered 3:1 at an argon pressure of 0.8 mTorr using DC magnetrons (Kurt J. Lesker CMS-A), from elemental gold and hafnium targets of purity better than 99.99% and 99.9% respectively, in a cryo-pumped UHV process chamber with base pressure below $10^{-8}$ Torr. Three films were deposited at room temperature (RT), 132 °C, and at 350 °C respectively, onto pre-cleaned Si wafers coated with a 300 nm oxide layer, while two other films were deposited at 200 °C and 415 °C on D263 technical glass. By using the technical glass it allows identification of any changes at the substrate/film interface during annealing studies. No adhesion layers were applied between the substrates and the deposited films. The samples were in-situ annealed in the process chamber at 450 °C for 1 hour immediately after deposition, to remove thermal stress and promote grain growth. Comparative samples were deposited at RT and 350 °C without the in-situ anneal, onto pre-cleaned Si wafers coated with a 300 nm oxide layer. Structural properties were characterised by x-ray reflection (XRR), and x-ray diffraction (XRD) on a Bruker D8 Discover X-ray diffractometer (Cu K$_\alpha$, $\lambda$ = 1.5406 Å), determining film thickness and crystalline composition, respectively, and using Bruker Leptos and EVA software for data analysis.

$Au_3Hf$ manifests as an orthorhombic structure, belonging to the $Cu_3Ti$ prototype [25], and space group of Pmmn (59) with lattice parameters of a = 6.038 Å, b = 4.785 Å, and c = 4.877 Å [27]. The major diffraction peaks in the films were identified by peak fitting using these lattice parameters and space group.

The optical properties of the structures were determined by the use of a J. A. Woollam Co. Inc, M-2000VI spectroscopic ellipsometer operating within a spectral range from 370 nm to 1690 nm. To establish the thermal stability the films were annealed for a second time in a vacuum oven purged with argon gas. The temperature was raised to 500 °C, at 3 °C /min, and dwelled at this peak for 1 hour, with a chamber pressure of $1.5 \times 10^{-6}$ Torr.

III. Results and discussion

3.1 Structural properties

The thickness of the films was between 90 and 95 nm as measured by XRR. Figure 1 shows the XRD spectra (the intensity is normalised to the maximal peak in the range of 25° to 55°) of the films.

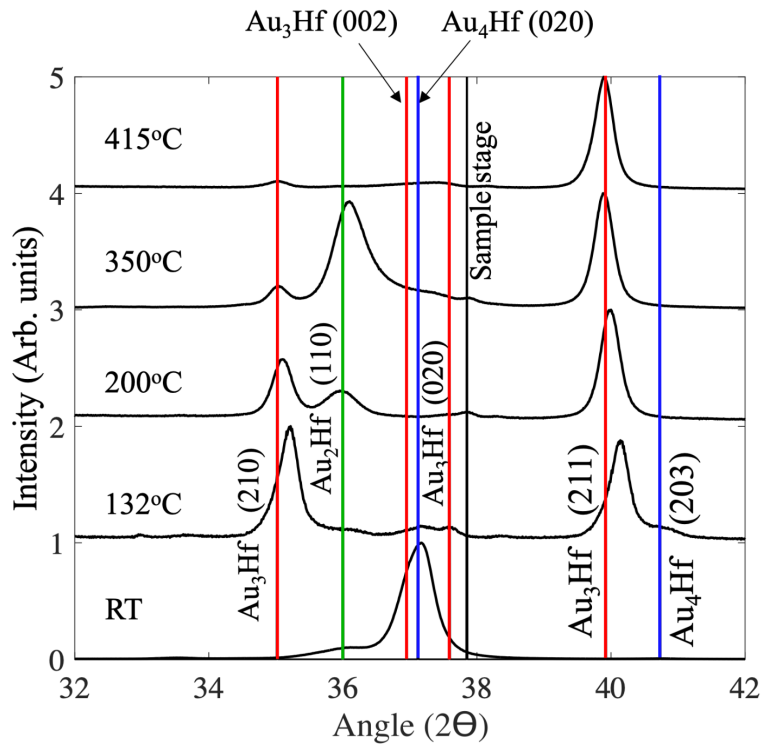

Fig.1 XRD spectra of Au$_3$Hf films fabricated at different deposition temperatures. Solid lines indicate the expected peak positions for Au$_3$Hf (red line), Au$_2$Hf (green line), Au$_4$Hf (blue line) calculated from the lattice parameters in literature, and the sample stage (black line).

From Fig. 1 it can be seen that the deposited films, in-situ annealed, are composed of a mix of phases and crystal orientations. The main orientations appear to be Au$_3$Hf (210), Au$_3$Hf (211) and a mix of Au$_4$Hf (020) and Au$_2$Hf (110). The 4:1 and 2:1 stoichiometries form in space groups Pnma (62) and I4/mmm (139) respectively [27]. There are no obvious trends clearly defining the correlation between the phase and crystal orientation and the deposition temperature. However, all films fabricated above 100 °C have a strong presence of $Au_3Hf$ (211) and presence of $Au_3Hf$ (210), and for the films made between 132 °C and 350 °C $Au_3Hf$ (210) peak decreases and $Au_2Hf$ (110) peak increases with temperature. The film deposited at RT is dominated by an asymmetric peak aligning with $Au_3Hf$ (002) and $Au_4Hf$ (020). The film fabricated at 415 °C is exclusively dominated by $Au_3Hf$ (211) orientation.

3.2 Optical properties

The optical properties of the films were characterised by spectroscopic ellipsometry and the results are displayed in Fig. 2.

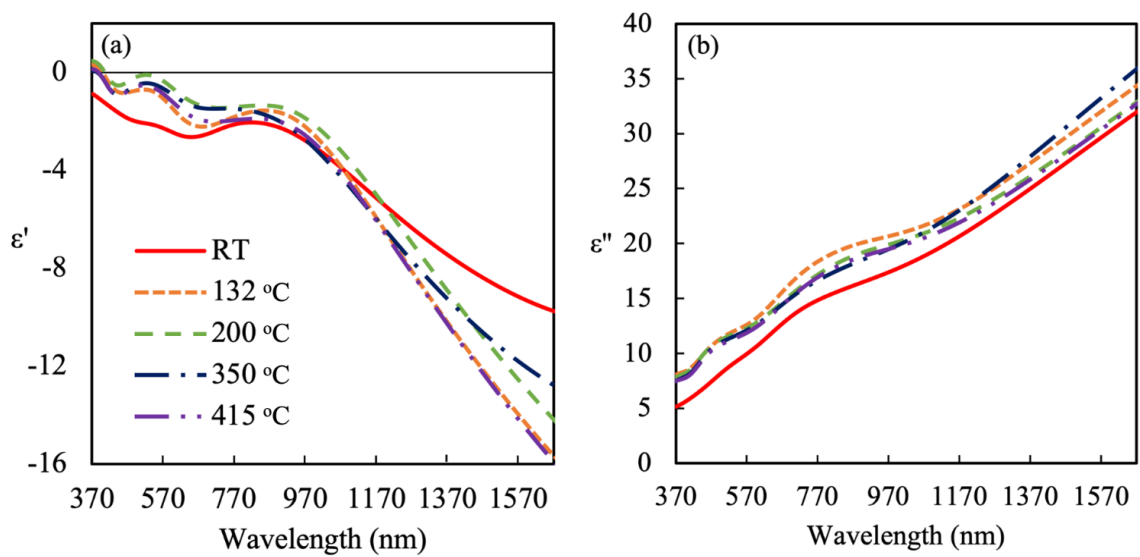

Fig. 2 Optical properties of the $Au_3Hf$ thin films fabricated at different deposition temperatures (°C). (a) Real ($\varepsilon'$) and (b) imaginary ($\varepsilon''$) parts of permittivity.

The optical results shown in Fig. 2 display clear plasmonic properties with the real components of permittivity (Fig. 2a) crossing to negative values at ~395 nm, aside from the RT film which intercepts at a lower wavelength outside the range of the figure. To further note, the optical properties appear approximately convergent for the majority of the films, with all the films made at 132 °C and above having very similar values with minimal divergence in the near infrared. With reference to Fig. 1, all of these convergent films have a dominant feature of $Au_3Hf$ (211) peak, which might suggest that this crystal orientation could play a dominant role in the resultant optical properties. However, crucially we find very similar optical properties of the 132 °C film and 415 °C film in the real permittivity, which might otherwise be expected to be one of the most disparate of the series, due to temperature difference. With reference to figure 1, the common feature between these films is the lack of a significant presence of the $Au_2Hf$ (110) phase, found in the 200 °C and 350 °C film. From this it may instead be suggested that the $Au_3Hf$ (210) phase, seen to decline with increasing deposition temperature, and $Au_3Hf$ (211) phases express similar optical properties, with the greater presence from the $Au_2Hf$ (110) phase in the other higher temperature deposited films causing the distinction.

The RT film is set apart from the other films. It has a smaller real part permittivity in the visible range and a larger real part permittivity in the near infrared range beyond 1000 nm, compared to the films fabricated at higher temperatures. The imaginary permittivity of the RT film exhibits a similar trend but is generally smaller than those of high temperature films. The notable deviation in the optical properties of RT film could be attributed to the Pnma crystal structure of the dominant $Au_4Hf$. As seen in Fig.1, the RT film is primarily dominated by $Au_4Hf$ (020), which is absent from the films fabricated at higher temperatures.

In Fig.3 we compare the optical properties of the RT and 415°C Au$_3$Hf films to those of equivalent Au$_3$Zr films [21], and find that the Au$_3$Hf films out-perform the Au$_3$Zr films, with more negative $\varepsilon'$ values across the full vis-NIR range for both films (Fig. 3a), and lower $\varepsilon''$ values across most of the spectrum up until ~1320 nm (Fig. 3b).

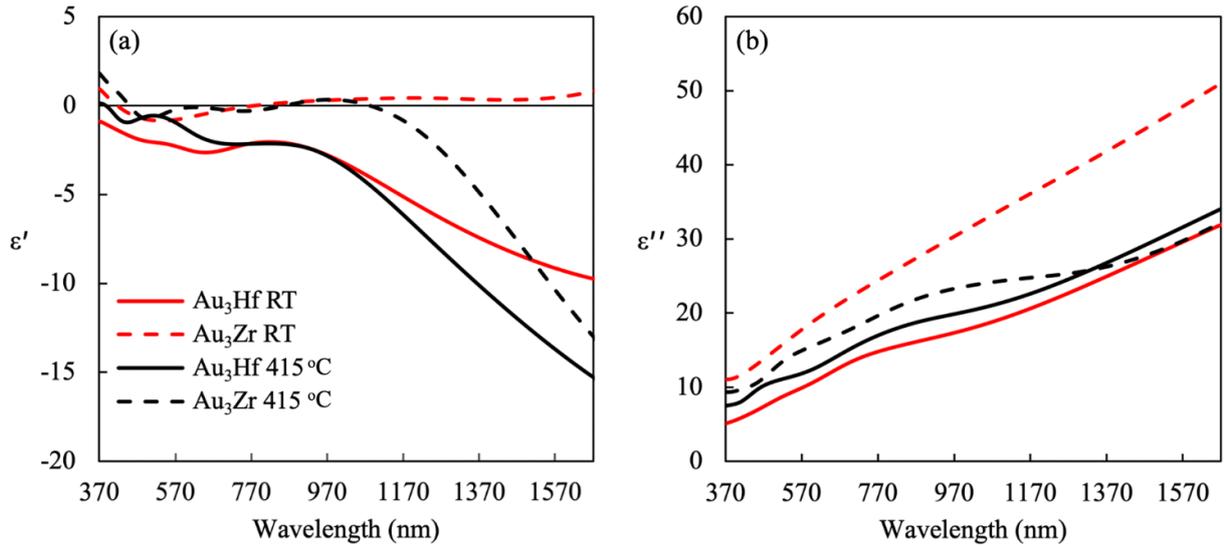

Fig. 3 The (a) real and (b) imaginary components of the permittivity of Au$_3$Hf (solid line) compared against Au$_3$Zr (dashed line) fabricated at RT (red lines) and 415 °C (black lines).

It is beneficial to obtain an empirical analytical expression of the optical properties of Au$_3$Hf films, which will be useful in theoretical modelling. For this we fitted the permittivity data with a Drude-Lorentz analytical model, as indicated in equation 1.

$$\varepsilon(\omega) = \varepsilon_\infty - \underbrace{\frac{\omega_p^2}{\omega(\omega + i\gamma)}}_{\text{Drude}} + \underbrace{\sum_i \frac{\omega_{p_i}^2}{\omega_{o_i}^2 - \omega^2 - i\gamma_i\omega}}_{\text{Lorentz}} \quad [1]$$

Where $\varepsilon_\infty$ is the permittivity at ultra-high frequency, $\omega_p$ is the plasma frequency associated with the Drude component, $\omega_{p_i}$ is the plasma frequency associated with the Lorentz component, $\omega$ is the frequency of oscillation, $\gamma_i$ is the damping factor, and $\omega_{o_i}$ is the eigenfrequency.

We used the Ref-fit software [31], which utilizes a Kramers-Kronig constrained variational fitting with a Levenberg-Marquardt algorithm to fit values. Given the similar spectra in Fig. 2, we used an average permittivity of the experimental data and fitted the model to that (the fitting values of each individual film are presented in Table S1 in the Supplementary Information). It is found that the experimental data can be well fitted with one Drude and two Lorentz oscillations. The results are presented in Fig. 4 with the fitting parameters shown in Table 1.

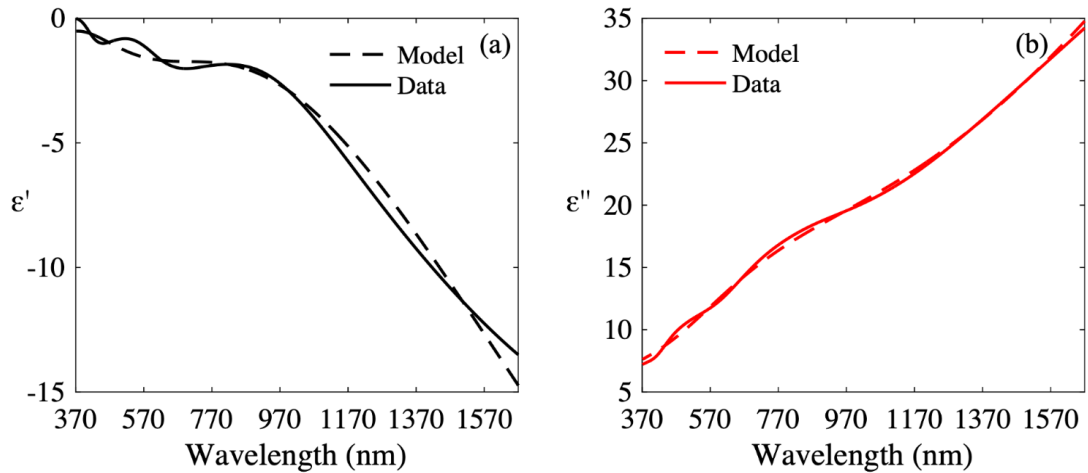

Fig. 4 The fitting of the analytical Drude-Lorentz model to the averaged permittivity data of Au$_3$Hf. (a) real and (b) imaginary permittivity.

As displayed in Fig. 4 the results of the fitting present a good approximation to the average permittivity of the films except the finer features of the data, including the wavy configuration in Fig. 4a between 370 and 870 nm, which can be better approximated with a greater number of Lorentz terms in the fitting.

Table 1 The fitting values of the Drude-Lorentz model to the averaged permittivity of the AuHf films from the combination of Drude (D) and two Lorentz ($L_n$) components in the analytical model. $\varepsilon_\infty$ is the permittivity at ultra-high frequency, $\omega_0$ is the eigen-frequency, $\omega_p$ is the plasma frequency, and $\gamma$ is the damping factor.

|  |  | $\omega_0$ [cm$^{-1}$] | $\omega_p$ [cm$^{-1}$] | $\gamma$ [cm$^{-1}$] |
|---|---|---|---|---|
| AuHf Average | D | – | 43648 | 4687.1 |
| $\varepsilon_\infty = 2.65$ | L1 | 14558 | 55882 | 20892 |
|  | L2 | 33956 | 65877 | 30601 |

Table 1 shows the results of the fitting parameters. From this we are able to determine the average plasma frequency of the Au$_3$Hf as 43648 cm$^{-1}$ or 8.22 ×10$^{15}$ rads$^{-1}$. When compared to gold with a value of 3.918 ×10$^{16}$ rads$^{-1}$ [32], it suggests a lower electron carrier density of Au$_3$Hf than Au. It is conjectured that inter/intraband transitions could be the basis of the contributions of the two Lorentzian oscillators in the model fit. These fitting parameters from the measurements of the optical properties will be useful for future theoretical modelling.

3.3 In-situ anneal effects in UHV chamber

We went on to investigate the effects of the in-situ anneal (in the process chamber of $1 \times 10^{-8}$ Torr at 450 °C for 1 hour immediately after deposition) on the structural and optical properties of the films. Fig. 5 compares the XRD spectra of the films deposited at 350 °C with and without in-situ anneal.

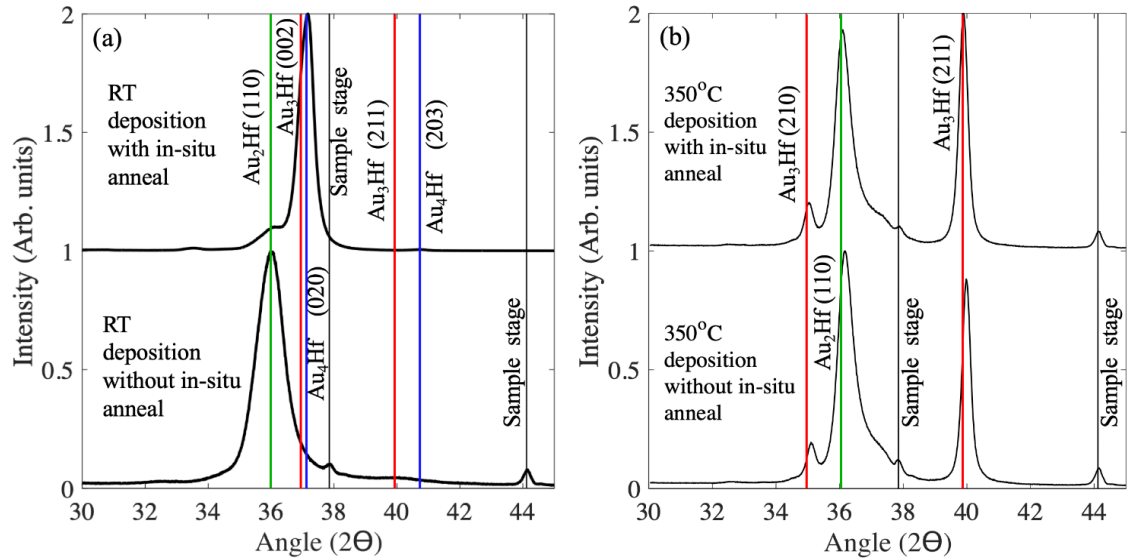

Fig. 5 Effects of in-situ anneal. Comparison of XRD spectra of Au₃Hf films deposited at (a) RT and (b) 350 °C, with and without in-situ anneal, respectively. The solid lines correspond to the expected peak positions for Au₃Hf (red line), Au₂Hf (green line), Au₄Hf (blue line), and the signal from the sample stage (black line).

In Fig. 5 (a) we compare the XRD spectra for the films made at RT with and without an in-situ anneal. There is a significant difference in the position of the most prominent peak, with the film without in-situ anneal dominated by a Au₂Hf (110) peak at 36.0°, whereas the film in-situ annealed finds a dominant peak at ~37.12°, which as more clearly seen in Fig. 1 aligns with a mix of Au₃Hf (002) and Au₄Hf (020). We also find that the width of the dominant peak with the in-situ anneal is smaller than that without the in-situ anneal, indicative of larger grains - a benefit expected from the in-situ anneal. For the films made at 350 °C (Fig. 5b), very little structural change was noted in the XRD spectra, with only fractional variance in the positioning of peaks. This suggests that films deposited at elevated temperatures are less susceptible to an in-situ anneal with respect to the structural properties.

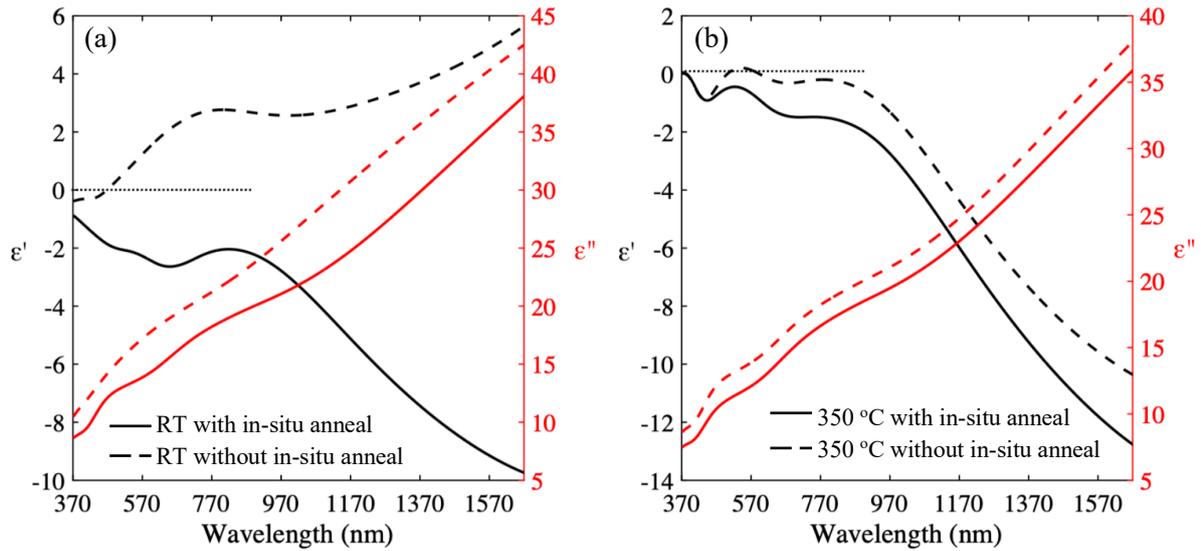

Fig. 6 The effects of in-situ anneal on the optical properties of $Au_3Hf$ films made at (a) RT and (b) 350 °C with (solid lines) and without (dashed lines) an in-situ anneal.

To determine what effects the in-situ anneal had on the optical properties, Fig. 6 compares the real and imaginary permittivity components for films made at RT and 350 °C, with and without in-situ anneal. It is evident from Fig. 6a that the difference in dominant phase and crystal orientation seen in Fig. 5a, results in a significant change in the optical properties of the AuHf films made at room temperature. When fabricated without in-situ anneal the film, dominated by the $Au_2Hf$ (110) orientation, has clearly dielectric properties, identified by the positive real component of the permittivity, however when made with in-situ anneal and with the dominant $Au_3Hf/Au_4Hf$ peaks we see strong indication of plasmonic properties with negative real component and a reduction in the imaginary component. This suggests that the $Au_2Hf$ (110) phase may be thermally unstable when deposited at RT, which shifts to the more stable $Au_3Hf$ (002)/$Au_4Hf$ (020) phases when in-situ annealed.

In Fig. 6b we compare the optical properties of the films made at 350 °C. With reference to the structural properties in Fig. 5b, we see minimal variation in the XRD spectra. This minimal variation translates to the optical properties presented, which are more or less similar with slight improvement in plasmonic properties seen as a reduction both in $\varepsilon'$ and $\varepsilon''$ values. This reduction in both $\varepsilon$ components is thought to be from a reduction in the damping within the film as a result of the anneal.

3.4 Oxidation effects

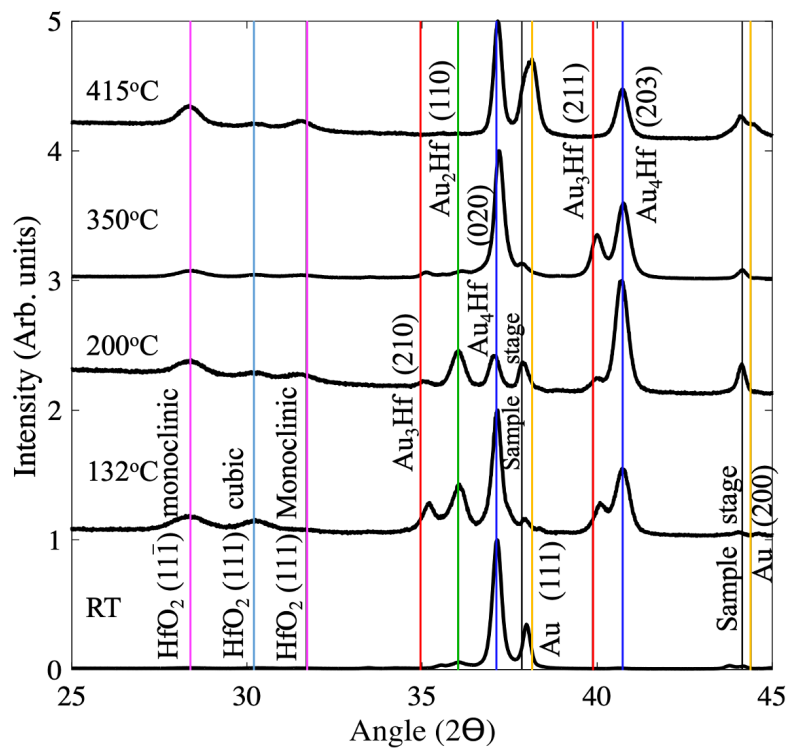

Fig. 7 XRD spectra of Au-Hf films annealed in a vacuum oven ($1.5\times10^{-6}$ Torr) at 500 °C for 1 hour. The solid lines correspond to the expected peak positions for $Au_3Hf$ (red line), $Au_2Hf$ (green line), $Au_4Hf$ (blue line), Au (gold line), $HfO_2$ monoclinic (magenta line) and $HfO_2$ cubic (grey line), and the signal from the sample stage (black line).

To further test the stability of the films, we annealed the films again in an oven of a lower vacuum level of $1.5×10^{-6}$ Torr at 500 °C for 1 hour, after the films had been annealed in-situ within the process chamber when fabricated. The XRD spectra of the films are shown in Fig. 7. A number of striking changes can be seen in these results with respect to Fig. 1, most prominently is the near disappearance of the $Au_3Hf$ phase and the appearance of the $Au_4Hf$ (020) and $Au_4Hf$ (203) peaks, which vary in relative proportions between films. Above 200 °C the $Au_4Hf$ (203) decreases with increasing temperature. A further change is the appearance of Au (111) and Au (200), most noticeably in the 415°C film (Au clusters can be seen on the sample surface under optical microscopy, with reference to Supplementary Information Fig. S1), in addition to three orientations of $HfO_2$ of two different (monoclinic and cubic) crystal structures. This suggests that the film surfaces have been partially oxidised with element/phase segregation by the trace oxygen in the oven. The films apparently most resistant to oxidation were the RT film and the 350°C film with weakly noticeable $HfO_2$ peaks. The disappearance of $Au_3Hf$ phase and the appearance of $Au_2Hf$ and especially the strong presence of $Au_4Hf$ suggests $Au_3Hf$ is less oxidation resistant at elevated temperature than other phases, and $Au_4Hf$ could be the thermally favourite phase for high temperature applications, which will be an interesting topic for future investigation.

Such structural changes would result in altered optical properties. However, the formation of surface crystallites on the samples, disrupting the homogeneity of the film surface, means accurate determination of the optical properties of the $Au_3Hf$ films post-anneal in the vacuum oven has not been possible. Films with the greater density of crystallites report negative $\varepsilon''$ values at short wavelength, an unphysical result.

IV. Discussion and conclusion

To put these results in the wider context, the average optical properties of the Au$_3$Hf thin films have been compared against known plasmonic materials, including the intermetallic AuAl$_2$; Al- doped and Ga-doped zinc oxide (AZO, GZO) and tin-doped indium-oxide (ITO) of the transparent conducting oxides (TCOs); HfN, TiN, and ZrN, and Au and Ag in figure 8.

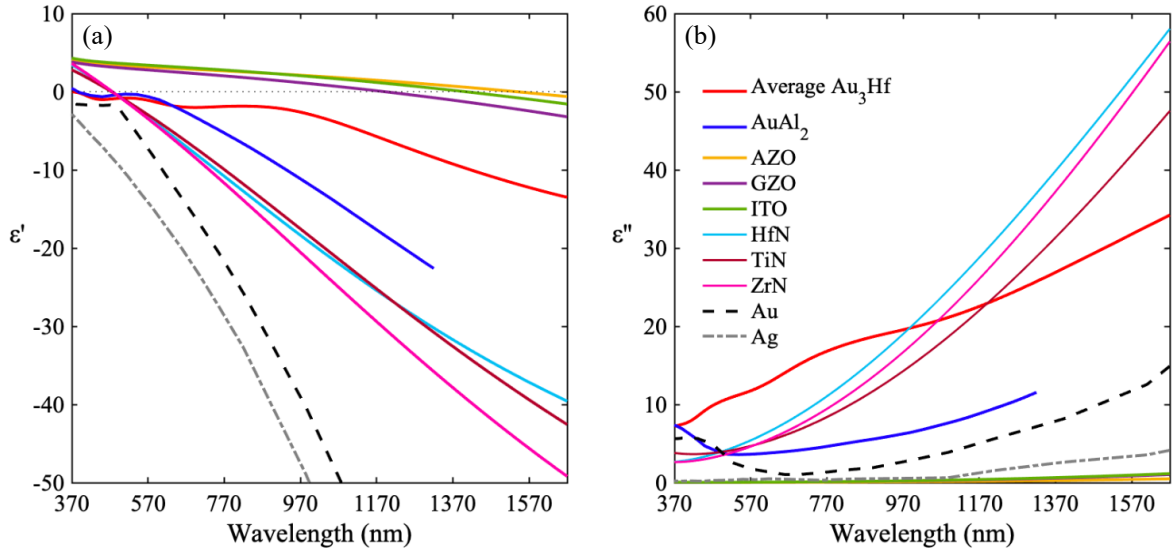

Fig. 8 The average (a) real and (b) imaginary permittivity of the Au$_3$Hf films discussed, compared against other plasmonic materials, including AuAl$_2$ [33], TCOs [34], nitrides [35] and Ag and Au [36].

With respect to figure 8 (a), Au$_3$Hf has very similar characteristics to AuAl$_2$ up until ~ 630 nm (approximately where the interband transitions of Au end), with negative $\varepsilon'$ values from the start of the wavelength range. Otherwise, it strikes an approximate mid-point of $\varepsilon'$ with the other materials, reaching a maximum of $\varepsilon' = -13.4$ at 1670 nm. Looking to figure 8 (b), the average Au$_3$Hf displays relatively high $\varepsilon''$ compared to the other materials between 370 – 970 nm, but lower than the metal-nitrides in the near infrared region beyond 1170 nm. While not exhibiting the metallicity of the nitrides in the infrared, the lower losses might make it an attractive candidate in the toolbox of alternate plasmonic materials.

In summary, here we for the first time fabricated and characterised the optical properties of Au$_3$Hf thin films. We have demonstrated that films of Au$_3$Hf can be made by DC magnetron co-sputtering of elemental Au and Hf at a ratio of 3:1. Below 400 °C a mix of Au$_3$Hf and Au$_2$Hf is found, with a possible inclusion of Au$_4$Hf at room temperature. The films are clearly seen to be plasmonic in the visible and near-infrared region, which appear predominantly influenced by the Au$_3$Hf (211) orientation when fabricated at above 100 °C. We fitted an analytical model for the averaged optical permittivity of the Au$_3$Hf films with good concordance, though more Lorentz terms may be needed to capture the finer features at shorter wavelengths. The films are thermally stable when annealed in an ultrahigh vacuum chamber with a base pressure of $10^{-8}$ Torr, but are partially oxidized when annealed at $1.5 \times 10^{-6}$ Torr at the expense of the Au$_3$Hf phase, changing composition to a mix of other phases. This represents the first experimental investigation into the optical properties of the Au$_3$Hf intermetallic, laying a foundation for further investigations on other Au-Hf alloys, such as Au$_4$Hf which shows evidence of better thermal stability and oxidation resistance at elevated temperature.

DECLARATIONS

FUNDING

This work was supported by the Engineering and Physical Sciences Research Council (Grant number EP/L015323/1).

RMB & WH acknowledge the support of Seagate Technology (Ireland) under SOW #00077300.0. RMB's contribution to project was supported by the Royal Academy of Engineering under the Research Chairs and

# Supplementary Information

## Optical properties of Au-Hf thin films

*Hugh Littlehailes, William R Hendren, Robert M Bowman, and Fumin Huang*[*]

*Queen's University Belfast, School of Mathematics and Physics, University Road, Belfast, BT7 1NN*

*Correspondence email: f.huang@qub.ac.uk

Table S1. The fitting values for the AuHf films from the combination of Drude (D) and two Lorentz ($L_n$) components in the analytical model. $\varepsilon_\infty$ is the permittivity at ultra-high frequency, $\omega_0$ is the eigen-frequency, $\omega_p$ is the plasma frequency, and $\gamma$ is the damping factor.

| Deposition Temperature [°C] | $\varepsilon_\infty$ | | $\omega_0$ [cm$^{-1}$] | $\omega_p$ [cm$^{-1}$] | $\gamma$ [cm$^{-1}$] |
|---|---|---|---|---|---|
| RT | 2.891 | D | _ | 41005 | 5510.6 |
| | | L1 | 13653 | 10742 | 5520.7 |
| | | L2 | 18264 | 73303 | 39244 |
| 132 | 2.513 | D | _ | 40746 | 2813.3 |
| | | L1 | 12872 | 10780 | 4048.8 |
| | | L2 | 22546 | 1.2466×10$^5$ | 75979 |
| 200 | 2.309 | D | _ | 43734 | 4348.8 |
| | | L1 | 13116 | 48249 | 18096 |
| | | L2 | 35537 | 94312 | 49165 |
| 350 | 2.630 | D | _ | 43889 | 4969.3 |
| | | L1 | 19922 | 87781 | 40623 |
| | | L2 | 30279 | 22039 | 2566.1 |
| 415 | 2.880 | D | _ | 43754 | 4198.7 |
| | | L1 | 13488 | 50523 | 18361 |
| | | L2 | 32300 | 73562 | 35629 |

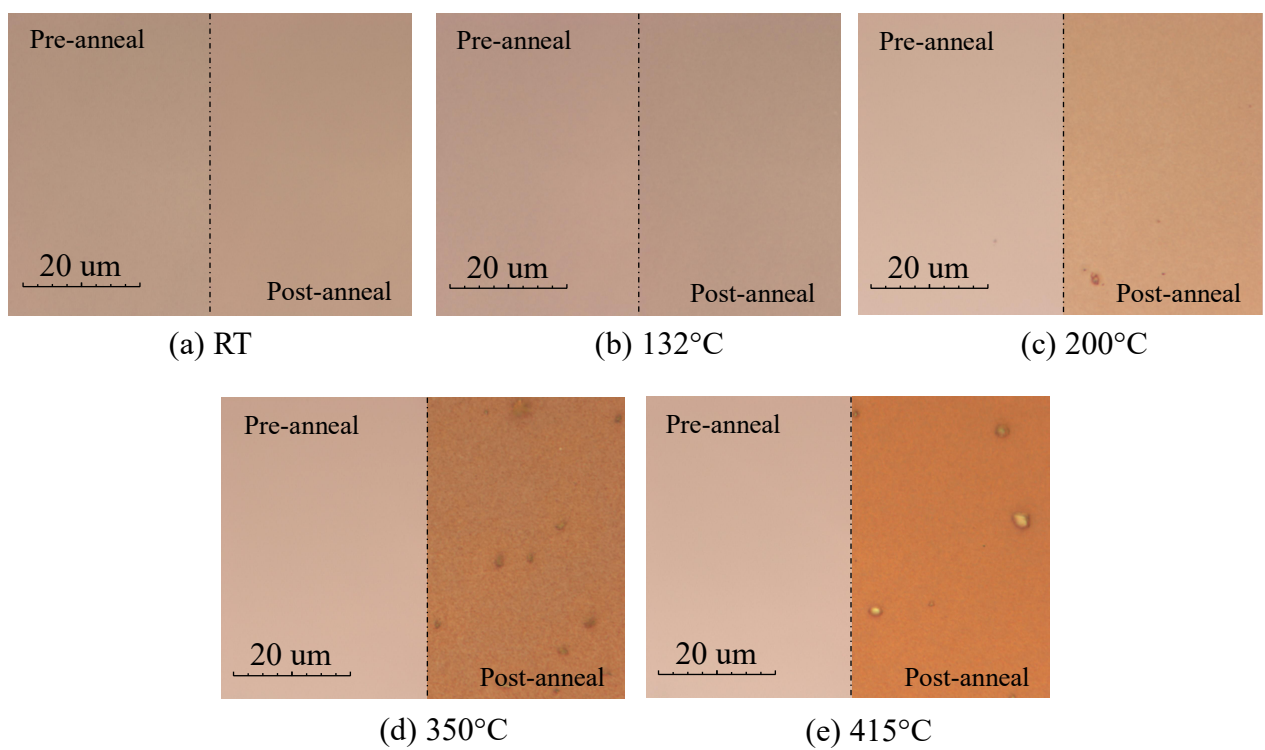

Fig. S1 Comparison of the optical images of Au$_3$Hf thin films before (Pre-anneal) and after (Post-anneal) being annealed in a vacuum oven of $1.5 \times 10^{-6}$ Torr at 497 °C for 1 hour. The temperature indicated is the deposition temperature of each film. Note that in addition to this ex-situ anneal, all films as part of the fabrication process had a 1 hour in-situ anneal at 450°C in the UHV conditions ($10^{-8}$ Torr) of the process chamber immediately following deposition. After the ex-situ, vacuum oven anneal, the Au$_3$Hf films were partly oxidized, decomposing into HfO$_2$ and Au which precipitate out and forms clusters that are clearly visible on the sample surface.